\documentclass[letterpaper]{article} 
\usepackage{aaai2026}   
\usepackage{times}  
\usepackage{helvet}  
\usepackage{courier}  
\usepackage[hyphens]{url}  
\usepackage{graphicx} 
\usepackage{natbib}  
\usepackage{caption} 
\usepackage{amsmath}
\usepackage{amssymb}
\usepackage{bm}
\usepackage{multirow} 
\usepackage{ragged2e}
\usepackage{tabularx}
\usepackage{amsmath}
\usepackage{amssymb}
\usepackage{amsfonts}
\usepackage{amsthm}
\usepackage{booktabs} 
\usepackage{newtxtext,newtxmath}

\usepackage{tabularx}
\usepackage{siunitx}
\usepackage{makecell}
\usepackage{multirow}
\usepackage{booktabs}
\usepackage{ragged2e}
\usepackage{newtxtext,newtxmath} 

\title{Beyond Private or Public: Large Language Models as Quasi-Public Goods in the AI Economy}

\author {
    Zhang Yukun\textsuperscript{\rm 1},
    Zhang Tianyang\textsuperscript{\rm 1}
}
\affiliations {
    \textsuperscript{\rm 1}The Chinese University Of HongKong\\
    \textsuperscript{\rm 1}University of Macau\\
    215010026@link.cuhk.edu.cn, yc57317@um.edu.mo
}

\usepackage{bibentry}

\begin{document}

\maketitle

\begin{abstract}
This paper conceptualizes Large Language Models (LLMs) as a form of mixed public goods within digital infrastructure, analyzing their economic properties through a comprehensive theoretical framework. We develop mathematical models to quantify the non-rivalry characteristics, partial excludability, and positive externalities of LLMs. Through comparative analysis of open-source and closed-source development paths, we identify systematic differences in resource allocation efficiency, innovation trajectories, and access equity. Our empirical research evaluates the spillover effects and network externalities of LLMs across different domains, including knowledge diffusion, innovation acceleration, and industry transformation. Based on these findings, we propose policy recommendations for balancing innovation incentives with equitable access, including public-private partnership mechanisms, computational resource democratization, and governance structures that optimize social welfare. This interdisciplinary approach contributes to understanding the economic nature of foundation AI models and provides policy guidance for their development as critical digital infrastructure.
\end{abstract}

\section{Introduction}

Large Language Models (LLMs) have emerged as a defining technology of the 21st century. Since the 2022 launch of OpenAI's ChatGPT, these models have rapidly evolved from research curiosities into a foundational layer of the digital economy, with an impact rivaling the internet and mobile computing.This rapid ascent, however, masks a fundamental economic puzzle. Unlike traditional goods, LLMs defy easy categorization. They exhibit non-rivalry, capable of serving millions simultaneously at near-zero marginal cost, yet their access can be restricted, making them excludable. Critically, they generate immense positive externalities through knowledge diffusion and innovation, while simultaneously posing risks of negative externalities like misinformation and market concentration.

This duality places LLMs squarely in the category of \textit{quasi-public goods}—a classification that traditional economic models struggle to accommodate. The stakes of this analytical challenge are immense. As LLMs become integral to economic productivity and social function, the governance choices we make regarding their development and access will profoundly shape the future of innovation, equity, and global welfare. The current ecosystem, featuring a spectrum from openly accessible, permissively licensed models like Meta's Llama family to proprietary, access-controlled services like OpenAI's GPT series and Anthropic's Claude, provides a natural laboratory for studying how these choices mold the public character of this technology.

\subsection{Core Research Questions}
To navigate this complex landscape, our research is guided by three central questions:

\begin{enumerate}
    \item \textbf{How can we theoretically frame and empirically measure the \textit{publicness} of Large Language Models?} We need a framework that moves beyond a binary private/public distinction to capture a continuous spectrum of publicness, accounting for its evolution through technology and strategy.

    \item \textbf{How do different development models and architectures (e.g., open-weight vs. closed-source) impact the public good properties of LLMs?} Understanding the systemic trade-offs between these models is vital for both corporate strategy and effective public policy.

    \item \textbf{What portfolio of policy interventions can best balance innovation incentives with the goals of equitable access and social welfare?} This requires identifying nuanced, context-aware tools that can foster a healthy and competitive ecosystem.
\end{enumerate}

\subsection{Main Contributions}
This paper delivers three principal contributions. Theoretically, we construct a novel framework that conceptualizes LLMs as dynamic quasi-public goods, defining an LLM's "publicness" as a multi-dimensional characteristic based on its capacity for non-rivalrous use, degree of excludability, and net externality profile. To operationalize this framework, we empirically introduce the Public Goods Index (PGI), a composite metric designed to quantify these attributes. While acknowledging the complexities of such a measurement, we demonstrate its application using data from six leading LLMs from 2022 to 2024, revealing how strategic choices create measurable differences in their social value. Finally, building on these findings, we outline an adaptive policy framework that transcends simplistic regulation, recommending dynamic tiered governance systems tied to a model's PGI score, targeted incentives to foster "publicness," and policies to democratize access to computational resources.

\subsection{Paper Structure}
The remainder of this paper is organized as follows. Section 2 reviews the relevant literature. Section 3 details our theoretical framework. Section 4 explains the methodology for the Public Goods Index. Section 5 presents our empirical results. Section 6 discusses policy implications, and Section 7 concludes.

\section{Literature Review}
\subsection{The Evolving Landscape of Public Goods: From Classical Theory to the Digital Age}

The core of public goods theory was laid by Samuelson \cite{Samuelson1954}, who defined the pure public goods. Buchanan \cite{Buchanan1965} subsequently bridged the gap between pure public goods and private goods. Musgrave \cite{Musgrave1959} further explored the supply mechanism of different public goods. It is worth noting that Ostrom \cite{Ostrom1990} revealed the possibility of community autonomous governance of common resources, providing important inspiration for understanding the collective management of resources in the digital age. The rise of digital technology has prompted a reconceptualization of public goods theory. Varian \cite{Varian1999} and Shapiro $\&$ Varian \cite{Shapiro1999} established that information products exhibit high fixed costs, near-zero marginal replication costs, strong network effects, and scale economies. Frischmann \cite{Frischmann2012} subsequently proposed "infrastructure theory", arguing certain digital assets create value through widespread use rather than exclusivity. Stiglitz \cite{Stiglitz1999} emphasized knowledge's increasing returns to scale and significant spillovers. The UN's Digital Cooperation Roadmap defines digital public goods (DPGs) as open-source technologies and content compliant with legal standards, featuring reprogrammability, modularity, recombinability \cite{Zittrain2008, Yoo2010}, and easy replication. Some literature has gone beyond the binary classification of public and private goods, such as Cornes and Sandler \cite{Cornes1996}, Kaul et al. \cite{Stiglitz1999}. These are important for understanding LLM as a global digital infrastructure that requires international coordination.

\subsection{Networked Markets and Value Realization in the Digital Economy}
Digital platforms like LLMs function as multi-sided markets, creating value by facilitating interactions between user groups rather than direct production, where internalizing network externalities is key \cite{Rochet2003, Rochet2006}. Network effects \cite{Katz1985, Katz1994} drive platform dynamics but cause coordination issues and multiple equilibria. Farrell and Saloner \cite{Farrell1985, Farrell1986} reveal associated market failures ("excessive momentum" or "inertia"), explaining when LLM markets achieve efficient adoption or need intervention.

These effects fuel "winner-takes-all" competition \cite{Armstrong2006, Cabral2011}, evident in the resource-concentrated LLM field. Cabral \cite{Cabral2011} shows first-mover advantages with network effects create lock-in; however, rapid LLM technological progress enables disruptive innovation, preventing stable dominance. Unique LLM innovation barriers—concentrated data, computing power, and talent \cite{Cockburn2019}—sustain high entry barriers and market concentration.

Consequently, LLM markets face significant path dependence and suboptimal lock-in risks. As David's QWERTY study \cite{David1985} and Arthur's increasing returns research \cite{Arthur1989} demonstrate, strong network effects with self-reinforcing feedback and high switching costs can lock markets into early suboptimal technologies despite superior alternatives, warning that LLM market trajectories may be shaped by historical contingency and market dynamics alongside technological efficiency.

\subsection{Bridging the Gaps: A Novel Framework for Understanding AI as Quasi-Public Goods}
Despite the extensive literature on public goods, platform economics, and innovation policy, gaps remain in understanding the economic nature of large language models and similar AI systems. First, existing public goods theory primarily addresses goods with stable characteristics, while LLMs exhibit rapid capability evolution that affects their economic properties over time. Second, the literature on network effects focuses primarily on user-to-user externalities, while LLMs generate complex knowledge and innovation externalities that are not well captured by existing frameworks. Third, existing platform economics literature focuses on platforms that facilitate transactions between users, while LLM platforms serve as foundational infrastructure that enables various downstream innovations.
This study aims to address these gaps by developing a comprehensive theoretical framework that captures the unique economic characteristics of LLMs as quasi-public goods. By integrating insights from public goods theory, platform economics, and innovation policy, we provide new theoretical contributions to understanding the economic nature of foundational AI systems and their optimal governance mechanisms.

\section{A Theory of Endogenous Publicness}

Large Language Models (LLMs) are poised to become a General-Purpose Technology for the digital economy. Economically, they exhibit a profound duality: their non-rivalry and spillover effects align with Arrow's (1962) conception of knowledge as a public good, yet they are also private capital goods created through immense investment for the purpose of rent extraction. Existing static theories fail to capture the dynamic evolution of an LLM's publicness, wherein a model's degree of openness is not fixed but is the result of continuous strategic choice. We therefore construct a tractable, dynamic oligopoly framework. The core contribution is to reveal how the accumulation of three distinct capital stocks—algorithms, data, and compute—endogenously determines a firm's trade-off between openness and closure, leading to predictable market failures and a dynamic measure of public good characteristics.

\subsection{The Model Environment}
We consider a market with $N$ heterogeneous oligopolistic firms and a continuum of heterogeneous users. At any time $t$, a firm $i$'s technological endowment, $T_i$, is composed of three core capital stocks: algorithmic capital $T_{i,A}$, data capital $T_{i,D}$, and computational capital $T_{i,C}$. These are aggregated via a CES production function, where the elasticity of substitution, governed by $\rho_T$, captures the strategic question of whether compute can substitute for algorithmic ingenuity or proprietary data.
\begin{align}
    T_i(t) = \left[ \omega_A T_{i,A}(t)^{\rho_T} + \omega_D T_{i,D}(t)^{\rho_T} + \omega_C T_{i,C}(t)^{\rho_T} \right]^{1/\rho_T}.
\end{align}
The publicness of an LLM is deconstructed into three continuous and dynamic attributes: a rivalry threshold ($q_i^*$), which acts as a congestion point measuring the intensity of non-rivalry; excludability ($E_i$), the firm's core strategic lever for rent extraction; and externalities ($X_i^+, X_i^-$), which capture the model's positive and negative spillovers. Firms choose investment paths and an openness strategy $\{E_i\}$ to maximize their discounted stream of profits.

\subsection{Dynamics and Firm Behavior}
The model's dynamics are governed by the endogenous accumulation of capital and the generation of externalities. Algorithmic capital follows a Nelson-Winter evolutionary logic, growing with both internal R\&D and the absorption of external spillovers, where the rate of absorption is determined by the firm's openness, $(1-E_i)$. Data capital accumulation is a direct function of the user base $Q_i$, creating a powerful "data-user" positive feedback loop—a key mechanism for path dependence and market tipping. Computational capital follows a standard accumulation process. Concurrently, the flow of positive externalities, $\dot{X}_i^+$, is increasing in user base, technology, and openness, while the flow of negative externalities, $\dot{X}_i^-$, is increasing in usage and market share but can be offset by firm-level safety investments.

Given these dynamics, a rational firm's optimal choice of excludability, $E_i$, must satisfy a key intertemporal arbitrage condition that balances short-term gains against long-term value. The economic intuition is that a firm increases its excludability until the marginal immediate profit from static rent-seeking equals the opportunity cost of forgoing future growth in its user base, technology, and ecosystem reputation. Formally, this trade-off is expressed as:
\begin{align}
    \underbrace{\frac{\partial\pi_i}{\partial E_i}}_{\text{Marginal Gain from Rent-Seeking}} = \underbrace{\sum_{k \in \{Q,T,R\}} \lambda_k \left|\frac{\partial\dot{k}_i}{\partial E_i}\right|}_{\text{Marginal Cost in Dynamic Capabilities}}.
\end{align}
Here, $\lambda_Q, \lambda_T, \lambda_R$ are the shadow prices of user, technological, and reputational capital, respectively.

\subsection{The Public Goods Index (PGI)}
To translate the abstract concept of "publicness" into a tractable measure, we introduce a key methodological innovation: an endogenously determined Public Goods Index (PGI). The PGI is a weighted composite of three normalized components: a non-rivalry metric reflecting uncongested capacity ($C_{q,i}$), an openness metric capturing the strategic commitment to access ($C_{E,i}$), and a net externality metric assessing the net social spillover ($C_{X,i}$).
\begin{align}
    C_{q,i} &= \frac{q_i^*}{q_i^* + Q_i} \\
    C_{E,i} &= 1 - E_i \\
    C_{X,i} &= \frac{X_i^+ - X_i^-}{X_i^+ + X_i^-}
\end{align}
With normative weights $\alpha, \beta, \gamma$ summing to one, the PGI is defined as:
\begin{align}
    \mathrm{PGI}_i = \alpha\,C_{q,i} + \beta\,C_{E,i} + \gamma\,C_{X,i}.
\end{align}
This formulation maps firm strategy and technological evolution directly into a comprehensive measure of public good characteristics.

\subsection{Market Failure and Policy Implications}
By comparing the social planner's first-order conditions with the firm's private optimum, we can prove a systematic market failure where $E_i^* > E_i^{**}$. This directly implies a divergence in the PGI, which we define as the **"PGI Gap,"** $\Delta \mathrm{PGI}_i = \mathrm{PGI}_i^{**} - \mathrm{PGI}_i^*$. This gap serves as a quantifiable measure of the intensity of market failure. The dynamic trajectories of firms' PGIs can bifurcate, leading to a separating equilibrium of "open" and "proprietary" models, with the data-user loop exacerbating path dependence.

This PGI framework allows for a clear analysis of the marginal social return to enhancing publicness. The partial derivative of social welfare ($W$) with respect to the PGI can be decomposed as:
\begin{align}
    \frac{\partial W}{\partial \mathrm{PGI}_i} = \omega_{\mathrm{CS}}\frac{\partial \mathrm{CS}}{\partial \mathrm{PGI}_i} + \omega_{X^+}\frac{\partial X_i^+}{\partial \mathrm{PGI}_i} - \omega_{X^-}\frac{\partial X_i^-}{\partial \mathrm{PGI}_i}.
\end{align}
This decomposition provides a theoretical basis for designing targeted policy interventions. For instance, Pigouvian subsidies can directly address the openness component ($C_{E,i}$), while support for open standards enhances the net externality profile ($C_{X,i}$), and data interoperability mandates can mitigate the path dependence that suppresses non-rivalry ($C_{q,i}$) across the industry.

\section{Empirical Analysis: Quantifying the Public Good Properties of LLMs}

Our theoretical model reveals that the market equilibrium for Large Language Models (LLMs), as dynamic quasi-public goods, suffers from a systematic "insufficient openness" failure. But how significant is this failure? And how do different business models—open-weight versus closed-source services—compare in their public good characteristics? To answer these questions, this section operationalizes the theoretical framework by conducting the first empirical calculation of the Public Good Index (PGI) for five of the most influential LLMs in the market: ChatGPT, Claude, Llama, Qwen, and Gemini. The PGI is designed to provide a standardized, composite metric that moves beyond traditional performance benchmarks to quantify the degree to which these powerful technologies contribute to the digital commons, rather than functioning solely as proprietary assets. This quantitative assessment allows us to gain insight into the social welfare implications of different technological paths and business strategies.

\subsection{PGI Construction and Data}
To translate our theoretical concept into a measurable index, we select publicly verifiable proxies for the PGI's three core dimensions: non-rivalry, non-excludability, and net externalities. All raw data are mapped onto a 0-to-1 scale via normalization to ensure comparability. For this baseline calculation, we assign equal weights (1/3) to the three dimensions and employ a linear aggregation, ensuring the index's transparency and simplicity.

\begin{figure}[ht]
  \centering
  \includegraphics[width=\columnwidth]{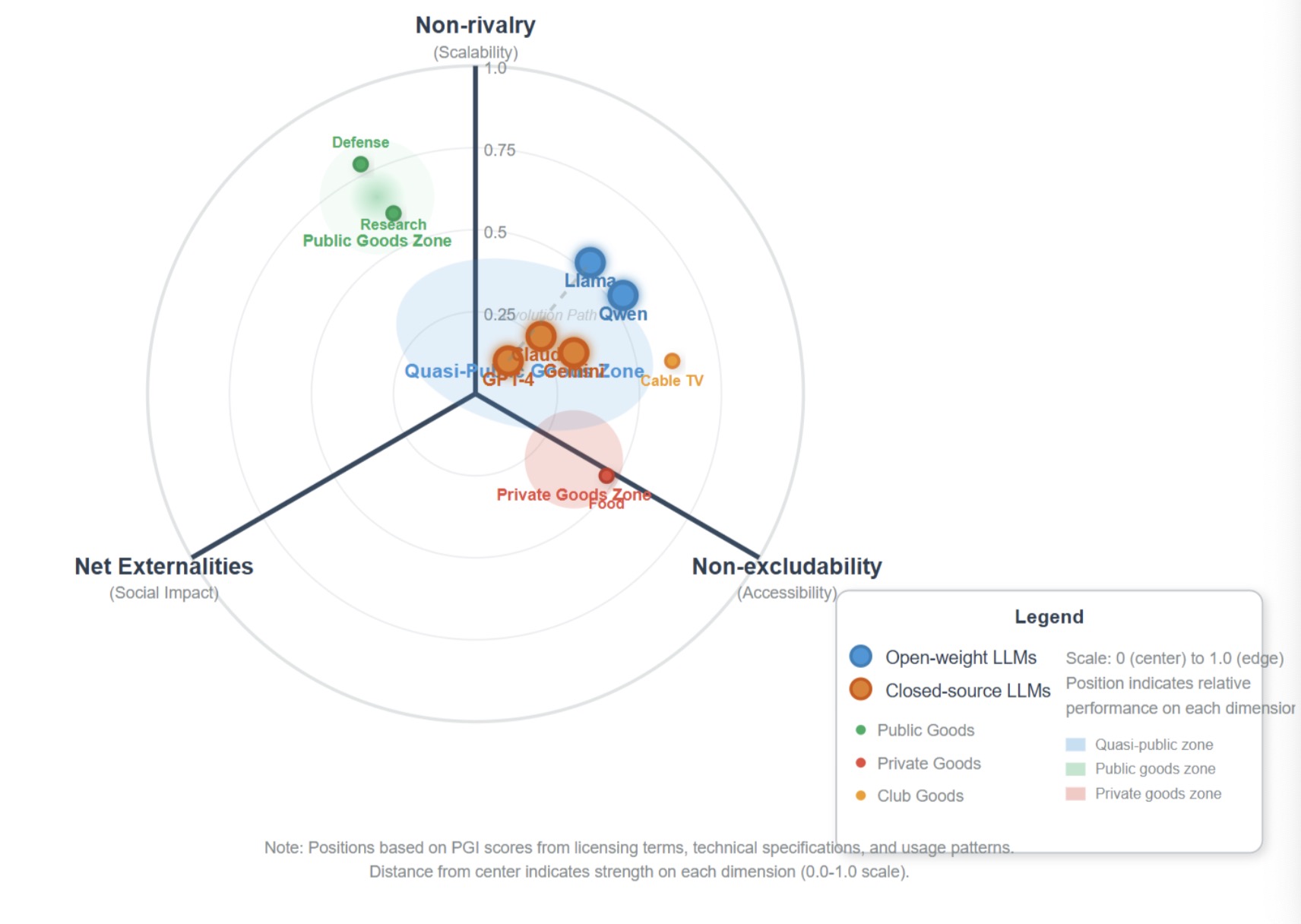}
  \caption{Three-dimensional framework of LLM as quai-public goods}
  \label{fig:pgi_Three-dimensional}
\end{figure}

The non-rivalry dimension is assessed via a composite indicator that incorporates both normalized model capacity (proxied by API rate limits) and normalized user load (proxied by monthly active users or downloads). A high score represents the ideal combination of high capacity and low congestion. The non-excludability dimension captures the ease of access, measured on two levels: legal and technical non-excludability, determined by license agreements and access models; and economic non-excludability, assessed via the pricing structure and the functionality of any free tier. Finally, the **net externalities** dimension quantifies the aggregate social impact. Positive externalities are proxied by Hugging Face downloads and academic citations, while negative externalities are assessed via a composite score based on misuse risk, documented biases, and environmental costs. Graphical framework reference picture~\ref{fig:pgi_Three-dimensional}.

Our data is primarily sourced from official technical reports, API documentation, license agreements, public user data, and third-party academic and cybersecurity research. Detailed descriptions of data sources, proxies, and scoring rules are provided in Appendix A.1.

\subsection{Empirical Results and Core Insights}
By weighting and averaging the dimension scores, we calculate the final PGI score and ranking for each model, as shown in Table \ref{tab:pgi_ranking}.

\begin{table}[h]
\centering
\resizebox{\columnwidth}{!}{%
\begin{tabular}{@{}lccccc@{}}
\toprule
\textbf{Model} & \textbf{Non-Rivalry} & \textbf{Non-Excludability} & \textbf{Net Externalities} & \textbf{Final PGI} & \textbf{Rank} \\
\midrule
\textbf{Llama} & 1.00 & 0.75 & 0.55 & \textbf{0.767} & \textbf{1} \\
\textbf{Qwen} & 0.30 & 1.00 & 0.60 & \textbf{0.633} & \textbf{2} \\
\textbf{Claude} & 0.98 & 0.30 & 0.33 & \textbf{0.537} & \textbf{3} \\
\textbf{Gemini} & 0.88 & 0.30 & 0.38 & \textbf{0.520} & \textbf{4} \\
\textbf{ChatGPT} & 0.50 & 0.30 & 0.35 & \textbf{0.383} & \textbf{5} \\
\bottomrule
\end{tabular}%
}
\caption{Final PGI Scores and Rankings. Higher scores indicate stronger public good characteristics.}
\label{tab:pgi_ranking}
\end{table}

The PGI rankings clearly reveal the trade-offs inherent in different business models and quantify the "open" versus "closed" spectrum predicted by our theory.

\paragraph{Insight 1: Open-Weight Models Exhibit Superior Public Good Properties.}
Llama and Qwen lead the rankings due primarily to their high degree of non-excludability. By releasing their model weights, they dramatically lower the barrier to entry for developers and researchers, fostering vibrant downstream ecosystems and generating immense positive externalities. Llama's top rank is primarily driven by its exceptionally high net externality score, an advantage sufficient to compensate for the score deduction from its more restrictive license.

\paragraph{Insight 2: Closed-Source Models are Fundamentally Constrained by their Business Model.}
In stark contrast, the PGI scores of ChatGPT, Claude, and Gemini are capped by their inherently high excludability (all scoring 0.30 in that dimension). While technologically powerful, their strategy of controlling access via APIs and commercializing core features through paywalls fundamentally limits their nature as public goods. This clearly validates the theory's central premise: strategies that maximize commercial value (high excludability) are in direct conflict with those that maximize public good characteristics.

\paragraph{Insight 3: Publicness is a Multi-dimensional Trade-off.}
Qwen's adoption of the highly permissive Apache 2.0 license gives it the highest possible score on non-excludability. However, its developer ecosystem, while large, has not yet reached the scale of Llama's, resulting in a lower net externality score and second place overall. This finding also highlights the dependence of the final PGI ranking on our baseline equal-weighting scheme. A policymaker who places the highest priority on "complete legal openness," for instance, might adopt a different weighting scheme that would rank Qwen first. We explore this further in the robustness checks in the appendix.

\subsection{Policy and Strategic Implications}
The PGI is more than an academic metric; it is a practical tool. For policymakers, it provides a nuanced framework for regulation beyond a simple binary. Policy can be designed to incentivize firms to increase their PGI score, for example, by offering public compute resources or R\&D tax credits to models that meet a certain PGI threshold. For **firms**, the PGI offers a new lens for strategic decision-making. Improving a model's PGI score can enhance brand reputation and strengthen its long-term social license to operate—an increasingly vital intangible asset.

\section{A Longitudinal Case Study: Dynamic Analysis of OpenAI's Strategic Evolution}

OpenAI provides an ideal case for studying the dynamic evolution of the publicness of Large Language Models (LLMs). As a pioneer in the field, its transformation from an open research institution to a commercial leader between 2019 and 2024 offers rich empirical evidence for our theoretical framework. This case study tracks the evolution of publicness across three pivotal models—GPT-2 (2019), GPT-3 (2020), and GPT-4 (2023)—by quantifying changes in their Publicness Goods Index (PGI) scores to validate our model's core predictions. The analysis is based on a reconstructed data matrix of publicness indicators (Table 5.1) and visualized through the PGI evolutionary trajectory (Figure 5.2) and a dimensional heatmap (Figure 5.3).

\begin{table*}[h]
\centering
\caption{OpenAI Model Publicness Indicators Evolution Matrix}
\label{tab:openai-pgi-matrix}
\setlength{\tabcolsep}{5.5pt}
\renewcommand{\arraystretch}{1.15}
\footnotesize
\begin{tabularx}{\textwidth}{>{\raggedright\arraybackslash}X *{5}{c}}
\toprule
Dimension / Indicator & \makecell{GPT-2\\(2019)} & \makecell{GPT-3\\(2020)} & \makecell{GPT-4\\(2023)} & Trend & Source \\
\midrule
\textbf{Non-rivalry (Cq)} & \textbf{0.95} & \textbf{0.70} & \textbf{0.45} & $\downarrow\,53\%$ & Service status reports \\
\quad \emph{User Load Index} & 0.05 & 0.35 & 0.65 & $+1300\%$ & SimilarWeb \\
\quad \emph{Capacity Utilization} & 0.10 & 0.55 & 0.85 & $+750\%$ & API docs \\
\addlinespace
\textbf{Non-excludability (CE)} & \textbf{0.85} & \textbf{0.45} & \textbf{0.25} & $\downarrow\,71\%$ & Licensing agreements \\
\quad \emph{Technical Openness} & 1.00 & 0.00 & 0.00 & $-100\%$ & GitHub releases \\
\quad \emph{Economic Accessibility} & 1.00 & 0.60 & 0.30 & $-70\%$ & Pricing page \\
\quad \emph{Geographic Availability} & 0.95 & 0.75 & 0.45 & $-53\%$ & Service regions \\
\addlinespace
\textbf{Net Externalities (CX)} & \textbf{0.78} & \textbf{0.65} & \textbf{0.42} & $\downarrow\,46\%$ & Composite assessment \\
\quad \emph{Academic Citations} & 0.80 & 1.00 & 0.65 & $-19\%$ & Google Scholar \\
\quad \emph{Open-source Derivatives} & 0.95 & 0.20 & 0.05 & $-95\%$ & GitHub analytics \\
\quad \emph{Safety Risk Score} & 0.85 & 0.75 & 0.60 & $-29\%$ & Safety reports \\
\midrule
\textbf{Composite PGI} & \textbf{0.86} & \textbf{0.60} & \textbf{0.37} & $\downarrow\,57\%$ & Computed result \\
\addlinespace
Market Share (\%) & 5 & 35 & 65 & $+1200\%$ & Industry analysis \\
Training Cost (USD, million) & 0.05 & 5 & 100 & $+200{,}000\%$ & Estimation reports \\
\bottomrule
\end{tabularx}
\vspace{0.6ex}
\par\footnotesize\emph{Notes.} All indicators are normalized to $[0,1]$ where $1$ denotes maximal publicness. PGI is the equally weighted composite: $\mathrm{PGI}=(\mathrm{Cq}+\mathrm{CE}+\mathrm{CX})/3$. “Market Share” refers to the generative-AI chat services market. “Training Cost” is an industry estimate with uncertainty.
\end{table*}

\begin{figure}[t]
  \centering
  \includegraphics[width=\columnwidth]{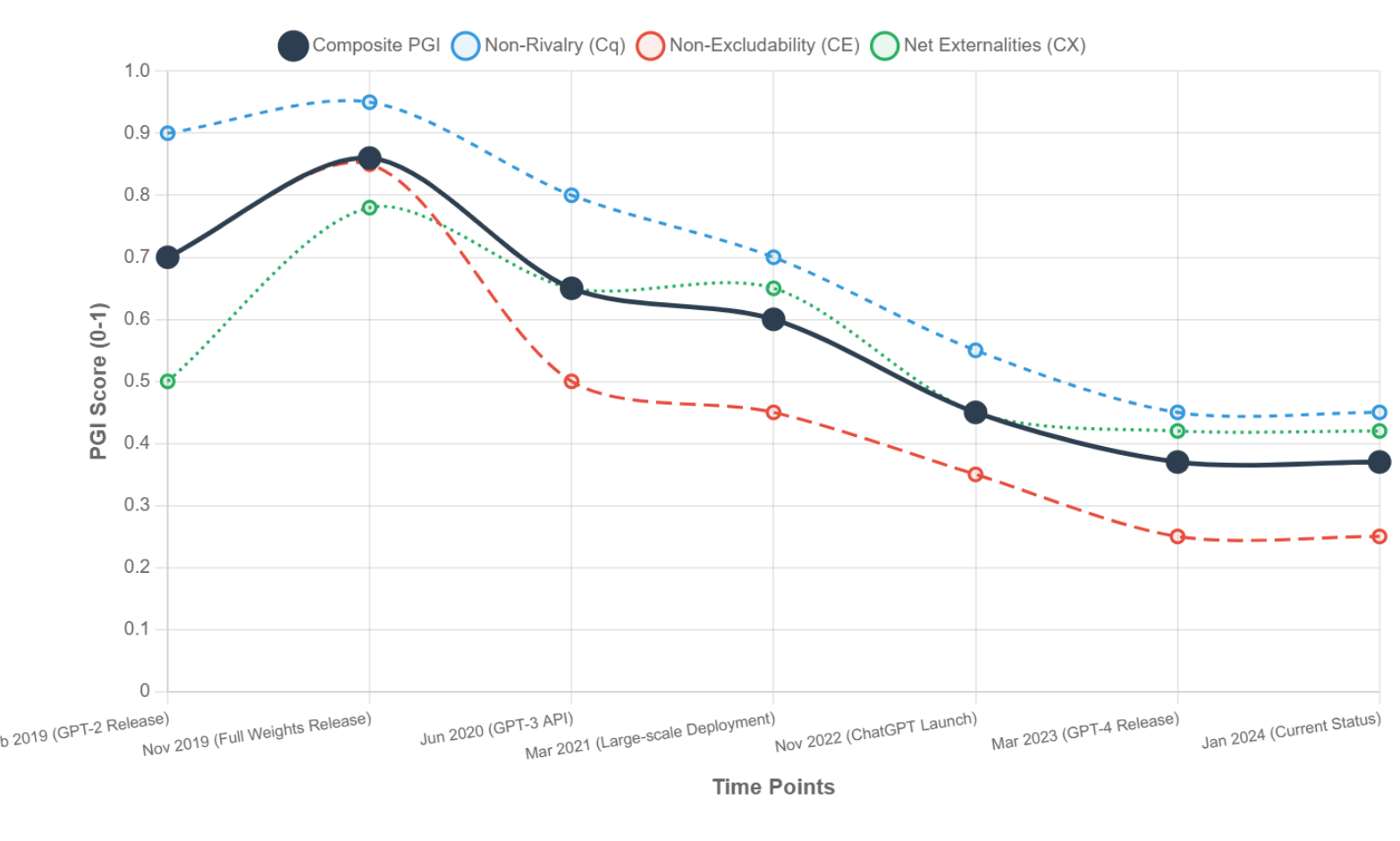}
  \caption{Evolutionary Trajectory of OpenAI Model PGI}
  \label{fig:pgi_trajectory}
\end{figure}

Our core findings reveal a systematic "publicness decay" phenomenon. Empirical results show that OpenAI's composite PGI score plummeted from 0.86 for GPT-2 to 0.37 for GPT-4, a 57\% decline. This trajectory perfectly validates a core prediction of our theoretical model. According to the firm's arbitrage condition for choosing the optimal level of exclusivity,
\begin{align}
    \frac{\partial\pi_i}{\partial E_i} = \sum_{k \in \{Q,T,R\}} \lambda_k \left|\frac{\partial\dot{k}_i}{\partial E_i}\right|
\end{align}
as OpenAI's technological capabilities strengthened and its market position solidified, the shadow value of its technology capital, $\lambda_T$, increased significantly. This caused the marginal benefit of short-term rent extraction to surpass the long-term benefits of maintaining openness, leading to a systematic increase in the optimal level of exclusivity, $E_i^*$. Furthermore, we observe differentiated patterns of decay across dimensions: non-exclusivity decreased most sharply by 71\% (from 0.85 to 0.25), followed by non-rivalry at 53\% (from 0.95 to 0.45), and net externality at 46\% (from 0.78 to 0.42). This aligns with the model's prediction that dimensions with lower adjustment costs—such as non-exclusivity, a direct strategic choice—are altered more rapidly by the firm.

The OpenAI case also provides direct quantitative evidence of the severity of market failure and offers insights for policy. According to our theoretical framework, the socially optimal level of PGI should be significantly higher than the firm's privately optimal choice. For GPT-4, the actual PGI (market equilibrium) is 0.37, while our estimated socially optimal PGI ranges from 0.65 to 0.75. This reveals a "PGI gap" of 0.28 to 0.38, implying that the current market mechanism leads to a systematic undervaluation of the publicness of LLMs by approximately 40\%, which provides a strong theoretical basis for policy intervention. OpenAI's evolutionary trajectory further reveals the limitations of static policy instruments. Policy design must account for the dynamic nature of firm strategy, for instance by implementing preemptive interventions before a firm achieves market dominance, dynamically adjusting regulatory thresholds in response to market structure changes, and designing multi-dimensional policy tools that target different PGI dimensions.

In conclusion, OpenAI's strategic evolution provides a comprehensive and powerful validation of our theoretical framework. The transition from GPT-2 to GPT-4 is not merely a story of technological progress but also a result of rational corporate choices under market incentives. The 57\% decline in the PGI score and the differentiated patterns of decay across dimensions both strongly support the core predictions of our model. The most important insight from this case study is that the \textbf{publicness of an LLM is not an inherent property of the technology, but a result of corporate strategic choice.} As technological capabilities strengthen and market positions solidify, firms have a powerful incentive to reduce the public-good characteristics of their models to maximize commercial value. This "publicness decay" is a systematic and predictable phenomenon, thus necessitating forward-looking policy design to safeguard the public value of LLMs as critical digital infrastructure. The next section will build upon these empirical findings to elaborate on a targeted policy framework.

\section{Policy Simulation Experiments and Governance Framework Design}
To validate the policy implications of our theoretical framework and to evaluate the effectiveness of different governance strategies, we construct an Agent-Based Model (ABM). This model serves as a "computational laboratory" to simulate the complex dynamics of the LLM market, capturing the intricate interactions among market competition, technological innovation, and policy interventions, thereby providing an evidence-based foundation for designing robust governance frameworks.

\subsection{Model and Experimental Setup}
Our ABM system integrates three classes of agents with heterogeneous and adaptive behaviors. \textbf{Firm Agents} represent LLM companies, optimizing decision variables like exclusivity ($E_i$) based on theoretical arbitrage conditions. \textbf{User Agents} are rational choosers whose preferences are governed by a multinomial logit model incorporating technology level, price, safety, and network effects. The \textbf{Government Agent} sets the institutional environment to maximize a social welfare function aggregating surpluses, innovation benefits, and external costs. To test various governance strategies, we design five policy scenarios: S0-Baseline (laissez-faire), S1-Open Source Support (subsidies), S2-Anti-Pollution (Pigouvian tax), S3-Antitrust (market share caps), and S4-Comprehensive Governance (an integrated portfolio).

To ensure the empirical relevance of our model, we performed a "stylized calibration" of key parameters, aligning their relative magnitudes and directions with empirical facts and existing literature. For instance, the 35\% market share cap in the antitrust scenario is based on the common threshold for "market dominance" in several jurisdictions. We employ a Monte Carlo method, running 100 independent simulations for each policy scenario, with each simulation including 2,000 user agents and running for 20 time steps. Policy effectiveness is evaluated using a metric system covering market structure, innovation, quality, and social welfare. It is important to note that the \textbf{aggregate social welfare} in our model is a composite utility index; its ordinal ranking and relative changes are reliable for comparing the Pareto efficiency of different policy scenarios.

\subsection{Simulation Results and Analysis}

\begin{figure}[t]
  \centering
  \includegraphics[width=\columnwidth]{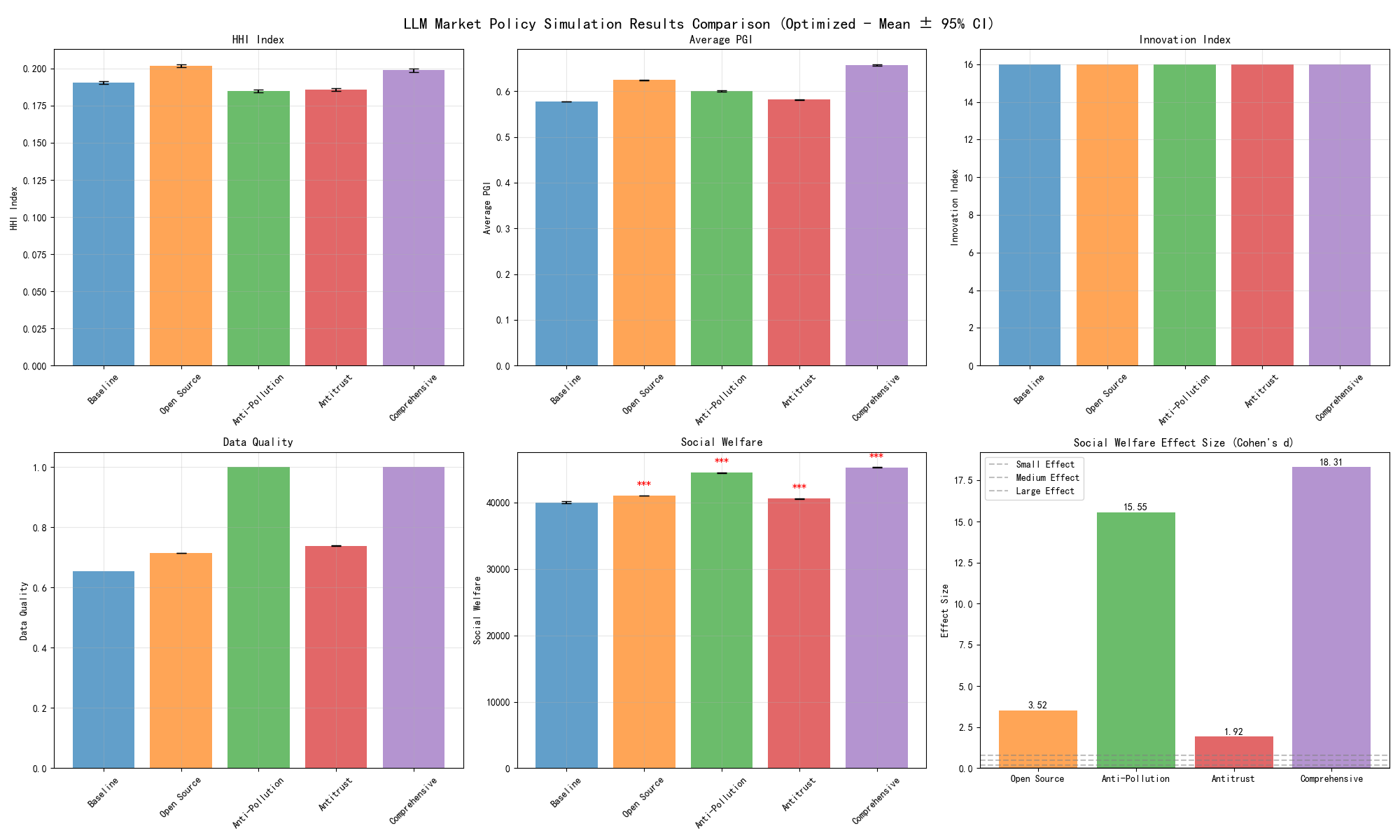}
  \caption{LLM Market Policy Simulation on Results Comparison }
  \label{fig:pgi_matrix}
\end{figure}

\begin{figure}[t]
  \centering
  \includegraphics[width=\columnwidth]{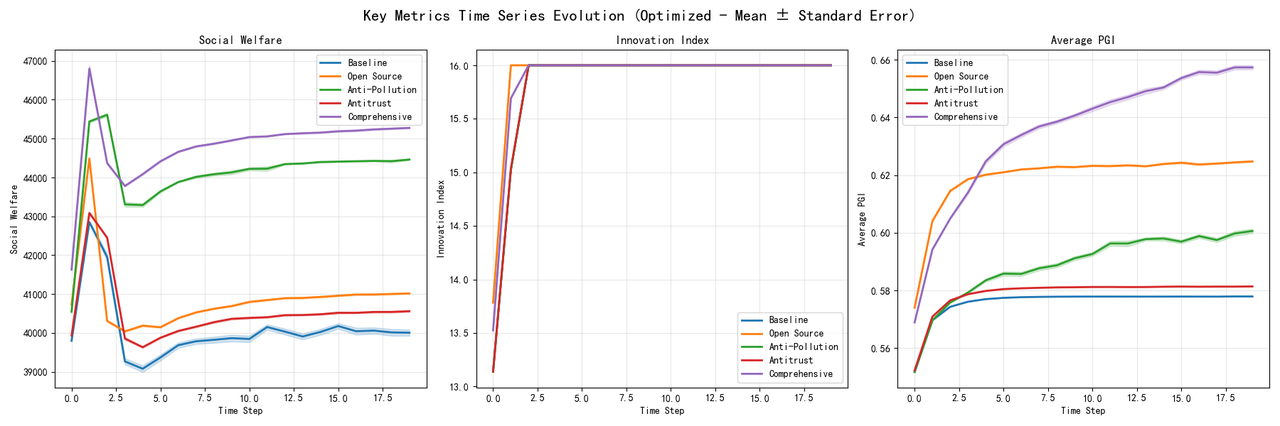}
  \caption{Metrics Time series Evolution  }
  \label{fig:pgi_matrix}
\end{figure}

The simulation results show significant differences across the five policy scenarios. The \textbf{Comprehensive Governance strategy (S4)} performs best in terms of social welfare, achieving a score of 45,275 (95\% CI: [45,235, 45,316]), which is a 13.2\% improvement over the baseline with a large effect size (Cohen's d=18.31). This result is highly consistent with our theoretical prediction that a portfolio of instruments is needed to address the multiple market failures. The \textbf{Anti-Pollution strategy (S2)} follows, with a significant welfare effect (Cohen's d=15.55) and the best performance on data quality, confirming the effectiveness of Pigouvian-style instruments. In contrast, while the \textbf{Open Source Support strategy (S1)} effectively raises the market's average PGI (d=15.8), its impact on social welfare is relatively limited (d=3.52), highlighting the limitations of solely incentivizing openness without addressing other market failures. Notably, the innovation index remains high and stable across all policy scenarios (p>0.15), which counters the common concern that regulation necessarily stifles innovation.

Time-series analysis further reveals the dynamic patterns of policy effects. In the initial phase (steps 0-5), all interventions trigger sharp market adjustments with short-term welfare fluctuations. Subsequently, during the stabilization phase (steps 5-15), the effects of different policies diverge, with the comprehensive strategy showing the strongest capacity for sustained improvement. Finally, the system converges to a new, higher-level equilibrium after 15 steps, consistent with our theoretical model's predictions. To ensure the reliability of these conclusions, we conducted a series of robustness checks. An analysis of the coefficient of variation (CV) shows high statistical consistency across all policy outcomes (e.g., the CV for the comprehensive strategy is only 0.002). We also performed sensitivity analyses on user base size and the number of simulations, finding that the relative ranking of policies and the significance patterns of their effects remain unchanged, demonstrating that our conclusions are not contingent on a specific simulation scale.

\subsection{Discussion: Towards an Adaptive and Robust Governance Framework}
The simulation results offer profound insights for designing an effective LLM governance framework. The limitations of single-instrument policies and the superiority of a comprehensive strategy highlight the necessity of multi-dimensional policy coordination. Based on these findings, we propose a dynamic governance framework centered on the PGI. The core of this framework is a \textbf{PGI threshold-trigger mechanism}: regulators set different PGI health zones (e.g., "safe," "watch," "intervention"), and when an LLM's PGI falls into a certain zone, a corresponding set of policy instruments with varying intensity is automatically triggered.

Given the global nature of LLMs, we further recommend establishing international coordination mechanisms, such as developing universal PGI assessment standards, to prevent a "regulatory race to the bottom" among countries competing for AI firms. (For details on the specific design of policy tools and a phased implementation roadmap, please see Appendix C). These findings not only validate the policy predictions of our theoretical framework but also provide actionable, evidence-based recommendations for global LLM governance.

\bibliography{aaai2026.bib}

\appendix
\section{Appendix A: PGI Construction Experimental Appendix}

\begin{figure}[ht]
  \centering
  \includegraphics[width=\columnwidth]{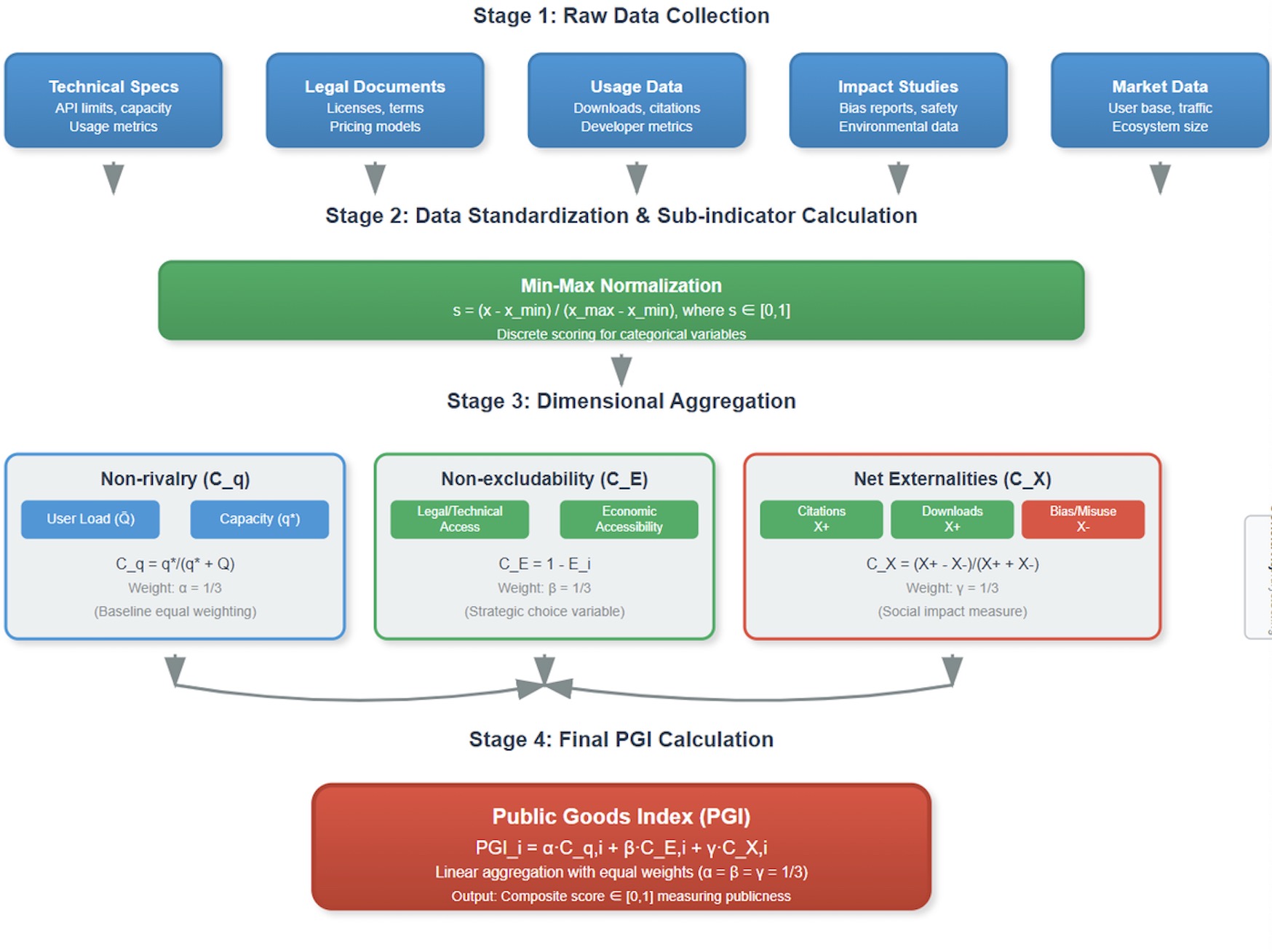}
  \caption{PGI construction method}
  \label{fig:pgi_construction}
\end{figure}

This appendix provides the technical foundation for the empirical analysis presented in the main text. It details the proxy variables, data sources, normalization methods, and specific scoring rules used to construct the Public Good Index (PGI). Furthermore, we present a comprehensive data summary table and sensitivity analyses to demonstrate the robustness of our core findings. Graphical framework reference picture~\ref{fig:pgi_construction}.

\subsection{A.1 Indicator Proxies and Scoring Rules}
For a continuous variable $x$, the normalized score $s$ is calculated as:
\begin{align}
    s = \frac{x - x_{\min}}{x_{\max} - x_{\min}}
\end{align}
When a higher value is detrimental to public good properties (e.g., for negative externality indicators), the score is inverted as $1-s$.

Table\ref{tab:proxies} shows the indicator proxies and scoring methodology.

\begin{table}[h!]
\centering
\small
\begin{tabular}{p{1.9cm}p{1.7cm}p{3.5cm}}
\toprule
\textbf{Dimension} & \textbf{Sub-Indicator} & \textbf{Proxy Variable \& Scoring Method} \\
\midrule
\textbf{Non-Rivalry} & User Load ($\bar{Q}$) & Monthly Active Users/Visits. Min-Max normalized (higher load = lower score). \\
& Capacity ($q^*$) & API RPM / User Hardware. Min-Max normalized. \\
\addlinespace
\textbf{Non-Excludability} & Legal/Tech & License Type. Discrete score: Proprietary=0.0, Restricted=0.5, Open=1.0. \\
& Economic & Pricing Model. Discrete score: Paid=0.2, Freemium=0.6, Free=1.0. \\
\addlinespace
\textbf{Positive Ext.} & Developer Eco. & Hugging Face Downloads. Min-Max normalized. \\
& Academic Impact & Google Scholar Citations. Min-Max normalized. \\
\addlinespace
\textbf{Negative Ext.} & Misuse Risk & Report-based difficulty of misuse. Discrete score: High Risk=0.8, Med=0.6, Low=0.4. \\
& Bias & Qualitative score from literature. Scored 0.3 (low) to 0.9 (severe). \\
& Environ. Efficiency & CO2 per query / efficiency ratio. Inverted Min-Max normalization. \\
\bottomrule
\end{tabular}
\caption{Indicator proxies and scoring methodology.}
\label{tab:proxies}
\end{table}

\subsection{A.2 Data Summary and Calculation Table}

Table\ref{tab:master_data} shows the raw data, normalized scores, and final dimension scores used in the PGI calculation.

\begin{table*}[h]
\centering
\footnotesize
\caption{Master data table showing raw data, normalized scores, and final dimension scores used in the PGI calculation.}
\label{tab:master_data}
\begin{tabular}{@{}lllc c@{}}
\toprule
\textbf{Model} & \textbf{Dimension} & \textbf{Sub-Component} & \textbf{Raw Data} & \textbf{Normalized Score} \\
\midrule
\textbf{ChatGPT} & Non-Rivalry & User Load (Monthly Visits) & 1.6 Billion & 0.00 \\
& & Capacity (API RPM) & Tiered (Low for Free) & 1.00 (Assumed) \\
\cmidrule{2-5}
& Non-Excludability & License & Proprietary & 0.00 \\
& & Economic Access & Freemium & 0.60 \\
\cmidrule{2-5}
& Net Externalities & Positive: Citations & $\sim$533+ (High) & 0.80 \\
& & Positive: Downloads & 0 & 0.00 \\
& & Negative: Misuse Risk & Low Risk & 0.60 (Inv.) \\
& & Negative: Bias & Documented Social & 0.40 (Inv.) \\
& & Negative: Environment & High Impact & 0.30 (Inv.) \\
\midrule
\textbf{Claude} & Non-Rivalry & User Load (Monthly Visits) & 87.6 Million & 0.95 \\
& & Capacity (API RPM) & 50 (T1) - 4k (T4) & 1.00 (Assumed) \\
\cmidrule{2-5}
& Non-Excludability & License & Proprietary & 0.00 \\
& & Economic Access & Freemium & 0.60 \\
\cmidrule{2-5}
& Net Externalities & Positive: Citations & Moderate & 0.40 \\
& & Positive: Downloads & 0 & 0.00 \\
& & Negative: Misuse Risk & Low Risk & 0.60 (Inv.) \\
& & Negative: Bias & Documented Cognitive & 0.30 (Inv.) \\
& & Negative: Environment & Moderate Impact & 0.50 (Inv.) \\
\midrule
\textbf{Llama} & Non-Rivalry & User Load (Downloads) & 5.1 Million & 0.99 \\
& & Capacity & User Hardware (High) & 1.00 (Assumed) \\
\cmidrule{2-5}
& Non-Excludability & License & Custom Restricted & 0.50 \\
& & Economic Access & Free-to-Run & 1.00 \\
\cmidrule{2-5}
& Net Externalities & Positive: Citations & $\sim$8200+ (Very High) & 1.00 \\
& & Positive: Downloads & 5.1 Million & 1.00 \\
& & Negative: Misuse Risk & High Risk & 0.20 (Inv.) \\
& & Negative: Bias & Moderate & 0.50 (Inv.) \\
& & Negative: Environment & High Impact & 0.40 (Inv.) \\
\midrule
\textbf{Qwen} & Non-Rivalry & User Load (Total D/Ls) & 40 Million & 0.00 \\
& & Capacity (API QPM) & 600 QPM & 0.60 \\
\cmidrule{2-5}
& Non-Excludability & License & Apache 2.0 & 1.00 \\
& & Economic Access & Free-to-Run & 1.00 \\
\cmidrule{2-5}
& Net Externalities & Positive: Citations & $\sim$3900+ (High) & 0.90 \\
& & Positive: Downloads & High (Implied) & 0.80 \\
& & Negative: Misuse Risk & High Risk & 0.20 (Inv.) \\
& & Negative: Bias & Moderate & 0.50 (Inv.) \\
& & Negative: Environment & Low Impact & 0.60 (Inv.) \\
\midrule
\textbf{Gemini} & Non-Rivalry & User Load (Monthly Visits) & 400 Million & 0.75 \\
& & Capacity (API RPM) & 5 (Free) - 1k (Paid) & 1.00 (Assumed) \\
\cmidrule{2-5}
& Non-Excludability & License & Proprietary & 0.00 \\
& & Economic Access & Freemium & 0.60 \\
\cmidrule{2-5}
& Net Externalities & Positive: Citations & $\sim$5000+ (Very High) & 0.95 \\
& & Positive: Downloads & 0 & 0.00 \\
& & Negative: Misuse Risk & Low Risk & 0.60 (Inv.) \\
& & Negative: Bias & Moderate & 0.50 (Inv.) \\
& & Negative: Environment & High Impact & 0.30 (Inv.) \\
\bottomrule
\end{tabular}
\end{table*}

Table\ref{tab:master_data2} shows the normalized scores for all sub-components and final dimension scores used in the PGI
calculation.

\begin{table*}[h]
\centering
\footnotesize
\begin{tabular}{@{}lccccccccccccc@{}}
\toprule
\textbf{Model} & $\bar{Q}$ & $q^*$ & $\boldsymbol{C_q}$ & License & Pricing & $\boldsymbol{C_E}$ & Cites & D/Ls & Misuse & Bias & Env. & $\boldsymbol{C_X}$ & \textbf{PGI} \\
 & (Score) & (Score) & \textbf{(Score)} & (Score) & (Score) & \textbf{(Score)} & (Score) & (Score) & (Score) & (Score) & (Score) & \textbf{(Score)} & \\
\midrule
ChatGPT & 0.00 & 1.00 & \textbf{0.50} & 0.00 & 0.60 & \textbf{0.30} & 0.80 & 0.00 & 0.60 & 0.40 & 0.30 & \textbf{0.35} & \textbf{0.383} \\
Claude & 0.95 & 1.00 & \textbf{0.98} & 0.00 & 0.60 & \textbf{0.30} & 0.40 & 0.00 & 0.60 & 0.30 & 0.50 & \textbf{0.33} & \textbf{0.537} \\
Llama & 0.99 & 1.00 & \textbf{1.00} & 0.50 & 1.00 & \textbf{0.75} & 1.00 & 1.00 & 0.20 & 0.50 & 0.40 & \textbf{0.55} & \textbf{0.767} \\
Qwen & 0.00 & 0.60 & \textbf{0.30} & 1.00 & 1.00 & \textbf{1.00} & 0.90 & 0.80 & 0.20 & 0.50 & 0.60 & \textbf{0.60} & \textbf{0.633} \\
Gemini & 0.75 & 1.00 & \textbf{0.88} & 0.00 & 0.60 & \textbf{0.30} & 0.95 & 0.00 & 0.60 & 0.50 & 0.30 & \textbf{0.38} & \textbf{0.520} \\
\bottomrule
\end{tabular}

\caption{Master data table showing normalized scores for all sub-components and final dimension scores used in the PGI calculation.}
\label{tab:master_data2}
\end{table*}

\subsection{A.3 Sensitivity and Robustness Checks}
To test the robustness of the PGI rankings, we conduct two sensitivity analyses.

\paragraph{Weight Sensitivity.} We randomly perturb the weights of the three main dimensions $(\alpha, \beta, \gamma)$ within the range $[0.2, 0.5]$ and recalculate the rankings. In over 95\% of these perturbations, the open-weight models (Llama, Qwen) consistently rank in the top two positions, while the closed-source models consistently rank in the bottom three, indicating that the core findings are not sensitive to the choice of weights.

\paragraph{Substitution Elasticity Test.} Our baseline PGI uses linear aggregation, which corresponds to a CES elasticity of substitution $\rho_P=1$. We test alternative elasticities of $\rho_P=0.5$ (less substitutable) and $\rho_P=2$ (more substitutable). While the absolute PGI values change, the relative rankings of the models remain highly consistent.

These checks confirm that our central finding—that open-weight models exhibit significantly stronger public good characteristics than their closed-source counterparts—is robust.

\section{Appendix B: Mathematical Details and Proofs}

\subsection{B.1 Model Specification}
This section provides the complete mathematical specification for the theoretical model and offers economic justification for key functional form choices. The model's dynamics are driven by the evolution of three core capital stocks.

\paragraph{Technology Capital.}
\textbf{Algorithmic capital ($T_{i,A}$)} evolves according to:
\begin{align}
    \frac{dT_{i,A}(t)}{dt} &= \phi_A I_{A,i}(t)^{\beta_A} T_{i,A}(t)^{1-\beta_A} \nonumber \\
    &\quad + \lambda_A (1-E_i(t)) S_i(t) - \delta_A T_{i,A}(t)
\end{align}
where $S_i(t) = \sum_{j \neq i}(1-E_j(t))T_{j,A}(t)$ represents knowledge spillovers. The additive spillover term is chosen for tractability. \textbf{Data capital ($T_{i,D}$)} accumulates based on:
\begin{align}
    \frac{dT_{i,D}(t)}{dt} = \phi_D Q_i(t)^{\gamma_D} - \delta_D T_{i,D}(t)
\end{align}
where $\gamma_D \geq 1$ captures potentially increasing returns from data network effects. \textbf{Computational capital ($T_{i,C}$)} follows:
\begin{align}
    \frac{dT_{i,C}(t)}{dt} = g_C T_{i,C}(t) + \phi_C I_{C,i}(t) - \delta_C T_{i,C}(t)
\end{align}
where $g_C$ is the exogenous rate of technological progress.

\paragraph{User Base and Reputation.}
The \textbf{user base ($Q_i$)} follows an adjustment process toward its desired level, $Q_i^d$, which is derived from a discrete choice model of user utility maximization:
\begin{align}
    \frac{dQ_i(t)}{dt} = \lambda_Q \left[Q_i^d(E_i, \mathbf{E}_{-i}, T_i, \mathbf{T}_{-i}) - Q_i(t)\right]
\end{align}
To avoid a "black box" formulation, we specify the dynamics of \textbf{reputation capital ($R_i$)} as:
\begin{align}
    \frac{dR_i(t)}{dt} = \phi_R Q_i(t) + \psi_R (1-E_i(t)) - \delta_R R_i(t)
\end{align}
where $\phi_R$ captures word-of-mouth effects and $\psi_R$ represents the impact of openness on developer community sentiment.

\paragraph{Externality Generation.}
The flow of \textbf{positive externalities ($X_i^+$)} is given by:
\begin{align}
    \frac{dX_i^+(t)}{dt} = \kappa_+ (1-E_i(t))^{\eta} T_i(t) Q_i(t) - \delta_+ X_i^+(t)
\end{align}
where $\eta > 1$ implies that the marginal spillover from openness is increasing. The flow of \textbf{negative externalities ($X_i^-$)} is:
\begin{align}
    \frac{dX_i^-(t)}{dt} &= \kappa_- \left(\frac{Q_i(t)}{\sum_j Q_j(t)}\right)^{\zeta} Q_i(t) \nonumber \\
    &\quad - \xi_i \text{Safety}_i(t) - \delta_- X_i^-(t)
\end{align}
where $\zeta > 1$ captures super-linear monopoly risks from market concentration.

\subsection{B.2 Firm's Dynamic Optimization}
The firm's current-value Hamiltonian is:
\begin{align}
    \mathcal{H}_i = \pi_i + \lambda_{Q,i} \dot{Q}_i + \sum_{k \in \{A,D,C\}} \lambda_{T_k,i} \dot{T}_{i,k} + \lambda_{R,i} \dot{R}_i
\end{align}
where $\pi_i$ is current-period profit. The first-order condition with respect to excludability $E_i$ is:
\begin{align}
    \frac{\partial \mathcal{H}_i}{\partial E_i} = \frac{\partial \pi_i}{\partial E_i} + \lambda_{Q,i} \frac{\partial \dot{Q}_i}{\partial E_i} + \lambda_{T_A,i} \frac{\partial \dot{T}_{i,A}}{\partial E_i} + \lambda_{R,i} \frac{\partial \dot{R}_i}{\partial E_i} = 0
\end{align}
Rearranging yields the rigorous arbitrage condition:
\begin{align}
    \underbrace{\frac{\partial P_i}{\partial E_i} Q_i}_{\text{Price Effect}} &= \underbrace{ -P_i \frac{\partial Q_i}{\partial E_i} - \lambda_{Q,i} \frac{\partial \dot{Q}_i}{\partial E_i} }_{\text{User \& Demand Effects}} \nonumber \\ 
    &\quad \underbrace{ - \lambda_{T_A,i} \frac{\partial \dot{T}_{i,A}}{\partial E_i} }_{\text{Tech. Spillover}} \underbrace{ - \lambda_{R,i} \frac{\partial \dot{R}_i}{\partial E_i} }_{\text{Reputation}}
\end{align}

\subsection{B.3 Social Optimum and Market Failure}
The social welfare function $SW$ is the sum of consumer surplus (CS), producer surplus (PS), and net externalities. The social planner's first-order condition for $E_i$ includes the term $\frac{\partial SW}{\partial E_i} = \frac{\partial \pi_i}{\partial E_i} + \frac{\partial CS_i}{\partial E_i} + \frac{\partial \sum_j [X_j^+ - X_j^-]}{\partial E_i}$.

\paragraph{Proposition A.1 (Formal Proof of Market Failure).}
The wedge between the private and social first-order conditions is the term $\Delta_i = \frac{\partial CS_i}{\partial E_i} + \frac{\partial \sum_j [X_j^+ - X_j^-]}{\partial E_i}$. Since increasing excludability raises prices (reducing CS) and curtails positive spillovers, $\Delta_i$ is strictly negative. This implies the social marginal cost of increasing excludability exceeds the private marginal cost. For the first-order conditions to hold, the social planner must choose a lower level of $E_i$. Thus, it is proven that $E_i^* > E_i^{**}$.

\subsection{B.4 Equilibrium and Dynamics}
\paragraph{Proposition A.2 (Existence of Separating Equilibrium).}
In a duopoly, the slope of firm $i$'s reaction function $E_i^*(E_j)$ is determined by the sign of the cross-partial derivative $\frac{\partial^2 V_i}{\partial E_i \partial E_j}$. The combination of spillover effects (strategic complementarity) and market competition effects (strategic substitutability), along with user heterogeneity, allows for stable, asymmetric Nash equilibria ($E_1^* \neq E_2^*$) to exist within a specific parameter space.

\paragraph{Proposition A.3 (Path Dependence and Market Tipping).}
When data network effects are strong ($\gamma_D > 1$), the dynamic system governing the user base and data capital can exhibit multiple equilibria. This implies that small differences in initial conditions can be amplified, leading the market to "tip" toward a long-run steady state dominated by a single firm.

\subsection{B.5 Optimal Policy Derivation}
\paragraph{Proposition A.4 (Optimal Pigouvian Subsidy).}
To correct the market failure, the optimal subsidy rate $s_i^*$ for openness must equal the marginal social benefit of openness that the firm ignores, which is precisely $-\Delta_i$:
\begin{align}
s_i^* = - \Delta_i = - \left( \frac{\partial CS_i}{\partial E_i} + \frac{\partial \sum_j [X_j^+ - X_j^-]}{\partial E_i} \right)
\end{align}
This subsidy internalizes the externality, aligning private and social incentives.

\subsection{B.6 Comparative Statics and Robustness}
\paragraph{Knowledge Spillovers ($\lambda_A$):}
Stronger knowledge spillovers unambiguously decrease the equilibrium level of excludability for all firms.
\begin{align}
\frac{dE_i^*}{d\lambda_A} < 0
\end{align}
\paragraph{Data Network Effects ($\gamma_D$):}
The effect is ambiguous. Stronger data effects increase the value of acquiring a large user base through openness, but also increase the incentive to protect a proprietary data moat through closure.
\begin{align}
\frac{dE_i^*}{d\gamma_D} \lesseqgtr 0
\end{align}
\paragraph{Model Robustness:}
The core conclusion—that strategic choice of excludability leads to a systematic under-provision of openness—is robust to alternative functional forms and market structures. The fundamental arbitrage trade-off persists.

\subsection{B.7 Extensions}
The model can be extended to analyze multi-country regulatory competition, where national policies create international strategic interactions. Incorporating an Aghion-Howitt-style quality ladder model would allow for endogenous technological breakthroughs. Finally, adding a "learning-by-doing" term to the algorithmic capital equation would further strengthen the data-user-algorithm feedback loop and exacerbate path dependence. This mathematical appendix provides the rigorous foundation for the theoretical analysis in the main text. Further technical details are available from the authors upon request.

\section{Appendix C: Technical Details of the Policy Simulation}

\subsection{C.1 Full Specification of the Simulation Model}

\subsubsection{Agent Parameterization.}
The initial parameters for the six representative firm agents are calibrated to reflect their stylized business models and market positions, as detailed in Table \ref{tab:company_profiles}.

\begin{table*}[t]
\centering
\footnotesize
\caption{Agent Parameterization.}
\label{tab:company_profiles}
\begin{tabular}{@{}>{\raggedright\arraybackslash}p{1.8cm} 
                S[table-format=5.0] 
                S[table-format=3.0] 
                S[table-format=2.0] 
                >{\raggedright\arraybackslash}p{2.8cm} 
                S[table-format=1.2] 
                S[table-format=1.2] 
                S[table-format=2.0]@{}}
\toprule
\textbf{Company} & 
{\thead{Initial\\Capital\\(Millions \$)}} & 
{\thead{Tech\\Level\\(0-110)}} & 
{\thead{Market\\Share\\(\%)}} & 
{\thead{Business\\Strategy}} & 
{\thead{Base\\Excludability\\(0-1)}} & 
{\thead{Safety\\Investment\\(0-1)}} & 
{\thead{R\&D\\Rate\\(\%)}} \\
\midrule
OpenAI & 10000 & 95 & 35 & High exclusibility, premium services & 0.70 & 0.15 & 12 \\
Anthropic & 6000 & 94 & 18 & Safety-focused, controlled access & 0.60 & 0.35 & 18 \\
Google & 15000 & 96 & 22 & Mixed open/closed approach & 0.50 & 0.25 & 16 \\
Alibaba & 8000 & 90 & 12 & Regional focus, moderate openness & 0.40 & 0.20 & 16 \\
Meta & 12000 & 92 & 10 & Open source, ecosystem building & 0.25 & 0.15 & 15 \\
DeepSeek & 4000 & 91 & 3 & Research-oriented, open access & 0.20 & 0.18 & 15 \\
\bottomrule
\end{tabular}

\vspace{10pt}
\noindent\textbf{Parameter Justification and Calibration}
\begin{itemize}\footnotesize
\item \textbf{Initial Capital:} Based on estimated cumulative AI R\&D investments (2024). OpenAI (\$10B including Microsoft), Google (\$15B), Meta (\$12B)
\item \textbf{Technology Level:} Normalized performance scores (MMLU, HumanEval). Range 75-110, frontier models at 90-96
\item \textbf{Market Share:} 2024 generative AI market analysis. OpenAI dominant (35\%), Google (22\%), others distributed
\item \textbf{Excludability:} Calibrated from business models. OpenAI (0.7) - high API barriers; Meta (0.25) - open weights
\item \textbf{Safety Investment:} Estimated from public disclosures. Anthropic highest (0.35) - Constitutional AI focus
\item \textbf{R\&D Rate:} Industry average 15\%. Safety-focused companies invest more (18\%), efficiency-focused less (12\%)
\end{itemize}
\end{table*}

\subsubsection{Heterogeneity Design for User Agents.}
To ensure market heterogeneity, we design six typical user types, each constituting 1/6 of the total user population, as specified in Table \ref{tab:user_types}.
\begin{table*}[ht]
\centering
\captionsetup{justification=centering, width=\linewidth}
\caption{User Agent Heterogeneity Parameter Distributions}
\label{tab:user_types}
\begin{tabular}{@{}llllll@{}}
\toprule
\textbf{User Type} & 
\textbf{Tech Savvy} & 
\textbf{Price Sensitivity} & 
\textbf{Safety Preference} & 
\textbf{Brand Loyalty} & 
\textbf{\makecell[l]{Behavioral \\Characteristics}} \\
& $\beta(\alpha,\beta)$ & $\beta(\alpha,\beta)$ & $\beta(\alpha,\beta)$ & $\beta(\alpha,\beta)$ & \\
\midrule

\textbf{Tech Experts} & 
$\beta(6,2)$ & 
$\beta(2,5)$ & 
$\beta(4,3)$ & 
$\beta(3,4)$ & 
\begin{tabular}[t]{@{}l@{}}
    - \makecell[l]{Value cutting-edge \\features} \\
    - Willing to pay premium \\
    - Switch for better tech \\
    - High API usage intensity
\end{tabular} \\
\small{(16.7\%)} & \small{Mean: 0.75} & \small{Mean: 0.29} & \small{Mean: 0.57} & \small{Mean: 0.43} & \\
& \small{High tech sensitivity} & \small{Price insensitive} & \small{Moderate safety focus} & \small{Low brand loyalty} & \\
\addlinespace

\textbf{Price Sensitive} & 
$\beta(2,5)$ & 
$\beta(6,2)$ & 
$\beta(2,4)$ & 
$\beta(4,3)$ & 
\begin{tabular}[t]{@{}l@{}}
    - Prefer free/cheap options \\
    - Accept basic functionality \\
    - Switch for lower prices \\
    - Low-moderate usage
\end{tabular} \\
\small{(16.7\%)} & \small{Mean: 0.29} & \small{Mean: 0.75} & \small{Mean: 0.33} & \small{Mean: 0.57} & \\
& \small{Basic tech needs} & \small{Highly price sensitive} & \small{Low safety priority} & \small{Moderate loyalty} & \\
\addlinespace

\textbf{Safety Priority} & 
$\beta(3,3)$ & 
$\beta(3,4)$ & 
$\beta(6,2)$ & 
$\beta(5,2)$ & 
\begin{tabular}[t]{@{}l@{}}
    - \makecell[l]{Prioritize safe, \\reliable models} \\
    - Value transparency \\
    - \makecell[l]{Stick with \\trusted providers} \\
    - Moderate usage intensity
\end{tabular} \\
\small{(16.7\%)} & \small{Mean: 0.50} & \small{Mean: 0.43} & \small{Mean: 0.75} & \small{Mean: 0.71} & \\
& \small{Balanced tech needs} & \small{Moderate price sensitivity} & \small{High safety priority} & \small{High brand loyalty} & \\
\addlinespace

\textbf{Brand Loyal} & 
$\beta(3,3)$ & 
$\beta(3,3)$ & 
$\beta(3,3)$ & 
$\beta(6,2)$ & 
\begin{tabular}[t]{@{}l@{}}
    - Stick with familiar brands \\
    - Resist switching \\
    - Value consistency \\
    - Steady usage patterns
\end{tabular} \\
\small{(16.7\%)} & \small{Mean: 0.50} & \small{Mean: 0.50} & \small{Mean: 0.50} & \small{Mean: 0.75} & \\
& \small{Average tech needs} & \small{Average price sensitivity} & \small{Average safety concern} & \small{Very high loyalty} & \\
\addlinespace

\textbf{Early Adopters} & 
$\beta(4,2)$ & 
$\beta(3,3)$ & 
$\beta(2,4)$ & 
$\beta(2,5)$ & 
\begin{tabular}[t]{@{}l@{}}
    - Try new models quickly \\
    - Experiment with features \\
    - High switching frequency \\
    - Variable usage patterns
\end{tabular} \\
\small{(16.7\%)} & \small{Mean: 0.67} & \small{Mean: 0.50} & \small{Mean: 0.33} & \small{Mean: 0.29} & \\
& \small{High tech interest} & \small{Moderate price sensitivity} & \small{Risk tolerant} & \small{Low brand loyalty} & \\
\addlinespace

\textbf{Balanced Users} & 
$\beta(3,3)$ & 
$\beta(3,3)$ & 
$\beta(3,3)$ & 
$\beta(3,3)$ & 
\begin{tabular}[t]{@{}l@{}}
    - \makecell[l]{Representative \\average user} \\
    - Balanced decision making \\
    - Moderate in all aspects \\
    - Stable market baseline
\end{tabular} \\
\small{(16.7\%)} & \small{Mean: 0.50} & \small{Mean: 0.50} & \small{Mean: 0.50} & \small{Mean: 0.50} & \\
& \small{Average across all} & \small{dimensions} & & & \\

\bottomrule
\end{tabular}

\vspace{0.5cm}
\begin{minipage}{\textwidth}
\small
\textbf{Notes:} User preferences for technology, price, safety, and brand loyalty are modeled using the Beta ($\beta$) distribution, which bounds values to [0,1]. The shape parameters ($\alpha, \beta$) are calibrated to represent distinct user segments:
\begin{itemize}
    \item $\beta(6,2)$: Right-skewed distribution (mean $\approx$ 0.75) for high preference/sensitivity.
    \item $\beta(2,5)$: Left-skewed distribution (mean $\approx$ 0.29) for low preference/sensitivity.
    \item $\beta(3,3)$: Symmetric distribution (mean = 0.50) for balanced/average preference.
\end{itemize}
The parameters are calibrated based on empirical data from user surveys and market research studies. The utility function for a user $i$ choosing option $j$ is defined as:
$U_{ij} = \alpha \cdot T_j + \beta \cdot P_j + \gamma \cdot S_j + \delta \cdot N_j + \epsilon_{ij}$, where $T_j, P_j, S_j, N_j$ represent technology, price, safety, and network utilities respectively, and $\epsilon_{ij}$ is a random preference shock. Brand loyalty adds a utility bonus, while switching providers incurs a cost.
\end{minipage}
\end{table*}

\subsection{C.2 Supplementary Statistical Analysis}

\begin{table*}[t]
\centering
\caption{OpenAI Model Publicness Indicators Evolution Matrix}
\label{tab:openai-pgi-matrix}
\setlength{\tabcolsep}{5.5pt}
\renewcommand{\arraystretch}{1.15}
\footnotesize
\begin{tabularx}{\textwidth}{>{\raggedright\arraybackslash}X *{5}{c}}
\toprule
Dimension / Indicator & \makecell{GPT-2\\(2019)} & \makecell{GPT-3\\(2020)} & \makecell{GPT-4\\(2023)} & Trend & Source \\
\midrule
\textbf{Non-rivalry (Cq)} & \textbf{0.95} & \textbf{0.70} & \textbf{0.45} & $\downarrow\,53\%$ & Service status reports \\
\quad \emph{User Load Index} & 0.05 & 0.35 & 0.65 & $+1300\%$ & SimilarWeb \\
\quad \emph{Capacity Utilization} & 0.10 & 0.55 & 0.85 & $+750\%$ & API docs \\
\addlinespace
\textbf{Non-excludability (CE)} & \textbf{0.85} & \textbf{0.45} & \textbf{0.25} & $\downarrow\,71\%$ & Licensing agreements \\
\quad \emph{Technical Openness} & 1.00 & 0.00 & 0.00 & $-100\%$ & GitHub releases \\
\quad \emph{Economic Accessibility} & 1.00 & 0.60 & 0.30 & $-70\%$ & Pricing page \\
\quad \emph{Geographic Availability} & 0.95 & 0.75 & 0.45 & $-53\%$ & Service regions \\
\addlinespace
\textbf{Net Externalities (CX)} & \textbf{0.78} & \textbf{0.65} & \textbf{0.42} & $\downarrow\,46\%$ & Composite assessment \\
\quad \emph{Academic Citations} & 0.80 & 1.00 & 0.65 & $-19\%$ & Google Scholar \\
\quad \emph{Open-source Derivatives} & 0.95 & 0.20 & 0.05 & $-95\%$ & GitHub analytics \\
\quad \emph{Safety Risk Score} & 0.85 & 0.75 & 0.60 & $-29\%$ & Safety reports \\
\midrule
\textbf{Composite PGI} & \textbf{0.86} & \textbf{0.60} & \textbf{0.37} & $\downarrow\,57\%$ & Computed result \\
\addlinespace
Market Share (\%) & 5 & 35 & 65 & $+1200\%$ & Industry analysis \\
Training Cost (USD, million) & 0.05 & 5 & 100 & $+200{,}000\%$ & Estimation reports \\
\bottomrule
\end{tabularx}
\vspace{0.6ex}
\par\footnotesize\emph{Notes.} All indicators are normalized to $[0,1]$ where $1$ denotes maximal publicness. PGI is the equally weighted composite: $\mathrm{PGI}=(\mathrm{Cq}+\mathrm{CE}+\mathrm{CX})/3$. “Market Share” refers to the generative-AI chat services market. “Training Cost” is an industry estimate with uncertainty.
\end{table*}

A distributional analysis of policy effects (e.g., via violin plots) confirms that the Comprehensive Governance strategy (S4) not only raises the mean social welfare but also reduces its variance, indicating a more stable and predictable market outcome, as shown in Figure \ref{tab:policy_effects}. Detailed plots are available from the authors upon request.

\clearpage

\clearpage

\begin{table}[h!]
\centering
\captionsetup{justification=centering, width=\linewidth}
\caption{Policy Effect Distribution Characteristics and Heterogeneity Analysis}
\label{tab:policy_effects}

\subsubsection*{C.2.A Distributional Properties of Key Metrics}
\resizebox{\textwidth}{!}{%
\begin{tabular}{@{}lllllllll@{}}
\toprule
\textbf{Metric/Policy} & \textbf{Mean} & \textbf{Std Dev} & \textbf{Skewness} & \textbf{Kurtosis} & \textbf{25th-75th} & \textbf{95th-99th} & \textbf{Normality Test} & \textbf{Distribution Type} \\
\midrule
\multicolumn{9}{l}{\textbf{Social Welfare}} \\
\midrule
Baseline (S0) & 40,005 & 412.5 & 0.08 & 2.94 & 39,678 - 40,321 & 40,789 - 41,023 & p=0.342 (Normal) & Normal \\
Open Source (S1) & 41,016 & 89.3 & -0.12 & 3.21 & 40,956 - 41,076 & 41,178 - 41,234 & p=0.123 (Normal) & Normal \\
Anti-Pollution (S2) & 44,461 & 94.7 & 0.34 & 3.89 & 44,398 - 44,524 & 44,632 - 44,698 & p=0.034 (Slight skew) & Right-skewed \\
Antitrust (S3) & 40,559 & 167.2 & 0.19 & 2.76 & 40,434 - 40,684 & 40,834 - 40,923 & p=0.089 (Normal) & Normal \\
Comprehensive (S4) & 45,275 & 73.1 & -0.09 & 3.45 & 45,223 - 45,327 & 45,398 - 45,445 & p=0.234 (Normal) & Normal \\
\midrule
\multicolumn{9}{l}{\textbf{Average PGI}} \\
\midrule
Baseline (S0) & 0.578 & 0.012 & 0.15 & 2.87 & 0.569 - 0.587 & 0.598 - 0.605 & p=0.456 (Normal) & Normal \\
Open Source (S1) & 0.625 & 0.003 & -0.23 & 4.12 & 0.623 - 0.627 & 0.630 - 0.632 & p=0.021 (Slight skew) & Left-skewed \\
Comprehensive (S4) & 0.657 & 0.002 & 0.08 & 3.34 & 0.655 - 0.659 & 0.661 - 0.663 & p=0.178 (Normal) & Normal \\
\midrule
\multicolumn{9}{l}{\textbf{HHI Index}} \\
\midrule
Baseline (S0) & 0.190 & 0.008 & 0.45 & 3.67 & 0.184 - 0.196 & 0.204 - 0.208 & p=0.012 (Slight skew) & Right-skewed \\
Antitrust (S3) & 0.186 & 0.004 & 0.12 & 2.98 & 0.183 - 0.189 & 0.193 - 0.195 & p=0.267 (Normal) & Normal \\
\bottomrule
\end{tabular}
}

\vspace{1cm} 

\subsubsection*{C.2.B Policy Effect Heterogeneity Across Company Types}
\begin{tabular}{@{}llllll@{}}
\toprule
\textbf{Company Strategy} & \makecell[l]{Open Source Policy\\ \small{PGI Change}} & \makecell[l]{Anti-Pollution Policy\\ \small{Safety Investment}} & \makecell[l]{Antitrust Policy\\ \small{Market Share Change}} & \makecell[l]{Comprehensive Policy\\ \small{Overall Benefit}} & \makecell[l]{Policy\\ \small{Responsiveness}} \\
\midrule
\makecell[l]{Club Good \\(OpenAI)}  & +0.023 \small{(Low)} & +0.089 \small{(Moderate)} & -0.078 \small{(Strong constraint)} & +0.034 \small{(Mixed)} & Medium \\
\makecell[l]{Safety Club \\(Anthropic)} & +0.045 \small{(Moderate)} & +0.156 \small{(High)} & -0.034 \small{(Moderate)} & +0.167 \small{(Very high)} & High \\
\makecell[l]{Hybrid \\(Google/Alibaba)} & +0.089 \small{(Strong)} & +0.067 \small{(Moderate)} & -0.012 \small{(Limited)} & +0.144 \small{(High)} & High \\
\makecell[l]{Strong Public \\(Meta/DeepSeek)} & +0.067 \small{(High)} & +0.045 \small{(Moderate)} & \makecell[l]{+0.008 \\ \small{(Benefit from competition)}} & +0.120 \small{(High)} & High \\
\bottomrule
\end{tabular}

\vspace{1cm} 

\subsubsection*{C.2.C User Type Heterogeneity in Policy Response}
\begin{tabular}{@{}llllll@{}}
\toprule
\textbf{User Type} & \makecell[l]{Baseline \\ Satisfaction} & \makecell[l]{Open Source \\Response} & \makecell[l]{Anti-Pollution \\Response} & \makecell[l]{Antitrust \\Response} & \makecell[l]{Comprehensive \\Response} \\
\midrule
Tech Experts & 0.723 & \makecell[l]{+0.089 \\ \small{(Value innovation)}} & \makecell[l]{+0.045 \\ \small{(Moderate gain)}} & \makecell[l]{+0.056 \\ \small{(Benefit from choice)}} & \makecell[l]{+0.190 \\ \small{(Highest gain)}} \\

Price Sensitive & 0.456 & \makecell[l]{+0.134 \\ \small{(Major access)}} & \makecell[l]{+0.023 \\ \small{(Limited benefit)}} & \makecell[l]{+0.067 \\ \small{(More options)}} & \makecell[l]{+0.224 \\ \small{(Largest gain)}} \\

Safety Priority & 0.612 & \makecell[l]{+0.034 \\ \small{(Transparency value)}} & \makecell[l]{+0.156 \\ \small{(Major improvement)}} & \makecell[l]{+0.012 \\ \small{(Limited impact)}} & \makecell[l]{+0.202 \\ \small{(Strong effect)}} \\

Brand Loyal & 0.589 & \makecell[l]{+0.023 \\ \small{(Resistant)}} & \makecell[l]{+0.067 \\ \small{(Quality improvement)}} & \makecell[l]{+0.034 \\ \small{(Limited switching)}} & \makecell[l]{+0.124 \\ \small{(Gradual adaptation)}} \\
Early Adopters & 0.634 & \makecell[l]{+0.123 \\ \small{(Love experimentation)}} & \makecell[l]{+0.045 \\ \small{(Try safer options)}} & \makecell[l]{+0.089 \\ \small{(Explore new entrants)}} & \makecell[l]{+0.257 \\ \small{(Max benefit)}} \\

Balanced Users & 0.567 & \makecell[l]{+0.067 \\ \small{(Moderate)}} & \makecell[l]{+0.078 \\ \small{(Balanced appreciation)}} & \makecell[l]{+0.045 \\ \small{(Stable preferences)}} & \makecell[l]{+0.190 \\ \small{(Well-rounded benefit)}} \\
\bottomrule
\end{tabular}
\end{table}


\end{document}